\documentclass[mathleft
]{an}
\usepackage{graphicx}
\usepackage{times}
\usepackage[british]{babel}
\begin{document}

\Pagespan{789}{}
\Yearpublication{2006}%
\Yearsubmission{2005}%
\Month{11}%
\Volume{999}%
\Issue{88}%

\title{The double-mode RR Lyrae variable BS Com}

\author{I. D\'ek\'any}

\titlerunning{The double-mode RR Lyrae variable BS Com}
\authorrunning{I. D\'ek\'any}
\institute{Konkoly Observatory, H-1525 Budapest, Pf.67, Hungary}

\received{}
\accepted{}
\publonline{}


\abstract{
We present the frequency analysis of the multicolour time series photometry of the field RRd variable BS Comae.
The large number of data points in each of the $BV(RI)_c$ bands and the $\sim0.01$ magnitude accuracy of the individual measurements allow us a high precision analysis of the properties of the combination frequencies due to nonlinear coupling.
Through the combination of the frequency spectra in different colors we show that except for the components corresponding to the linear combinations of the two pulsation modes, there are no other components present above the millimagnitude amplitude level.}

\maketitle

\section{Introduction}
Most of the RR Lyrae variables are known to pulsate either in the fundamental or in the first overtone mode (RRab and RRc stars, respectively). There are also a few evidences for the existence of second mode pulsators (RRe stars) in globular clusters (e.g. Walker \& Nemec 1996) and also in the Galactic field (Kiss et al. 1999) but these are still considered to be controversial (Smith 2006).
At a moderate fraction of RR Lyrae stars multiperiodicity can be observed. One of these is the phenomenon of periodic light curve modulation known as the Blazhko effect, which is known already for a century (see e.g. Jurcsik et al. 2005).
In the case of the other type (denoted as RRd), simultaneous excitation of the fundamental and first overtone radial modes take place.
In most cases the two types can be clearly distinguished observationally, since the period ratio of the latter type is well defined, and can vary only in a narrow interval around $\sim0.745$.\\
\indent Double-mode RR Lyrae stars are very important objects in astrophysics for various reasons.
In general, studying stellar oscillations is a good way to gain access to their physical properties and studying RRd's is perhaps the cleanest field of astroseismology because unambigous mode identification is possible only for radial pulsation.
In the case of RRd's we have a unique opportunity to estimate absolute physical parameters such as the stellar mass.
The schematic physics behind this is that radial oscillations mainly depend on the stellar mass and radius, and having two periods these parameters can be determined.
Furthermore, they are also important test objects of nonlinear stellar pulsation models in the case of two radial modes (see Koll{\'a}th et al. 2002).\\
\indent The first star identified as a double-mode RR Lyrae was AQ Leo by Jerzykiewicz \& Wenzel (1977). Since then a large number of RRd stars were discovered mostly in Galactic globular clusters and extragalactic objects (Magellanic Clouds and dwarf spheroidal galaxies). Unfortunately, only two dozens of them are known in the Galactic field -- a vast majority of these were found very recently by the analysis of the databases supplied by the automated surveys ASAS and NSVS (Wils 2006 and references therein). Although field RRd stars are of particular importance, yet most of them are very poorly studied. Because they are relatively bright, their spectroscopic follow-up study is possible with middle-sized telescopes. Additionally, since field RR Lyrae stars are distributed in a much wider metallicity interval than those found in clusters, they could be used to test the mass -- metallicity distribution on a much wider metallicity range.

\section{The data}
\sloppy{BS Com was first considered by us as a possible target for our ongoing photometric survey of Blazhko stars (see S\'odor in these proceedings) based on the investigations of Clementini et al. (2005). Therefore we put it on our observing schedule. In the meantime, its double-mode nature was published by Wils (2006), who made rough estimates to the periods by the analysis of ASAS-3 data.} We carried out four-color photometric observations with the 60 cm automatic telescope of the Konkoly Observatory, equipped with a Wright $750 \times 1100$ CCD, in the Johnson--Kron--Cousins $BV(RI)_C$ passbands. During the springs of 2005 and 2006 a total number of 35 nights of observation were made. Data reduction and relative aperture photometry were performed using stanard IRAF\footnote[1]{IRAF is distributed by the National Optical Astronomy Observatory,
which is operated by the Association of Universities for Research in Astronomy, Inc., under cooperative agreement with the National Science Foundation.} packages. Our large number of data points (almost 1000 in $V(RI)_C$ and nearly 600 in $B$) with good phase coverage constitutes the most complete RRd photometric data set to date, and enable us to perform a Fourier-analysis of high signal-to-noise ratio.

\begin{figure}[]
\includegraphics[angle=-90, width=83mm]{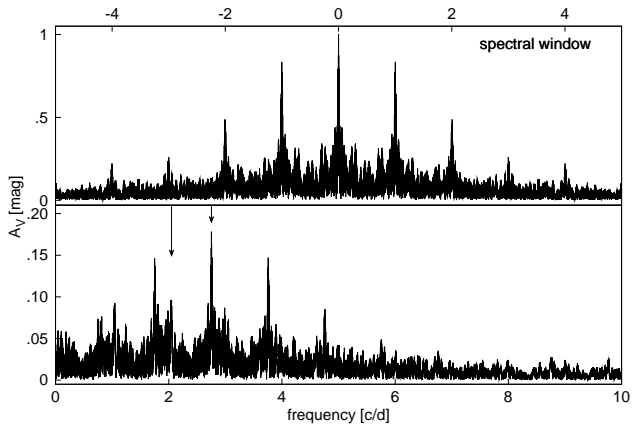}
\caption{ Spectral window \emph{(upper panel)} and amplitude spectrum \emph{(lower panel)} of the $V$ data of BS Com. The peaks corresponding to the radial fundamental and first overtone modes are marked with arrows.
}
\label{modes}
\end{figure}

\section{Frequency analysis}
\subsection{The $V$ light curve}
\begin{figure}
\includegraphics[angle=-90, width=83mm]{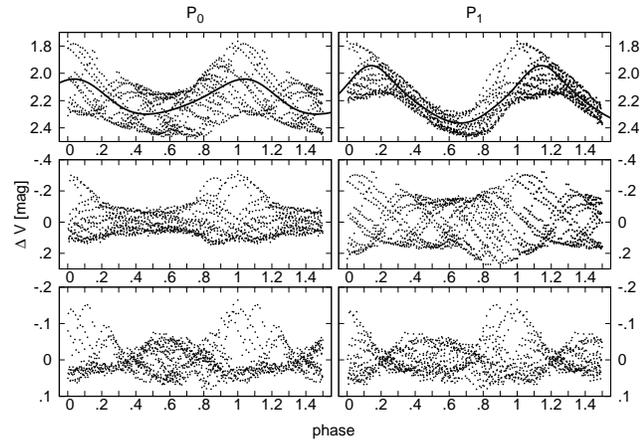}
\caption{$V$ light curve of BS Com folded with the periods of the fundamental and the first overtone mode (\emph{top panels}).
Synthetic light curves computed from the amplitudes and phases of the two radial modes and their harmonics are shown.
\emph{Middle panels} show the prewhitened fundamental and first overtone mode light curves. Strong variations remain showing the existence of additional high amplitude frequency components. In the \emph{bottom pa\-nels} residual light curves after prewithening the contibution of both modes are shown. The presence of other strong frequency components is evident, which are identified as the numerous linear combination frequencies in the Fourier spectrum. Their phase relations are clearly denoted: the residual light variation has the highest amplitude when both of the modes are around maximum brightness.
}
\label{foldedv}
\end{figure}

\begin{figure}
\includegraphics[angle=-90, width=83mm]{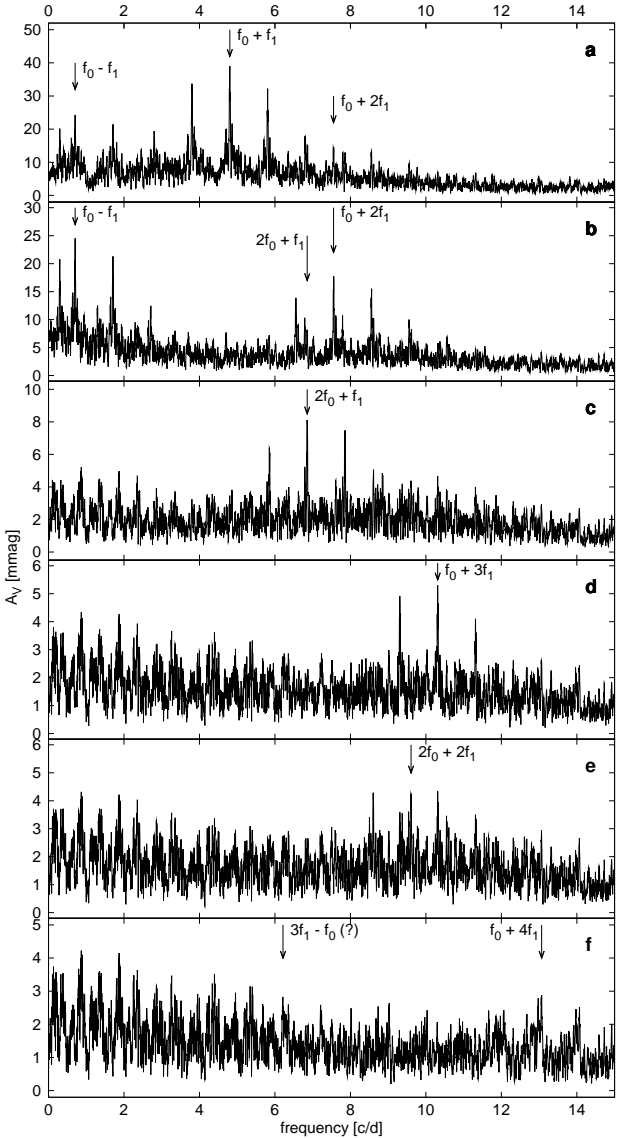}
\caption{Residual amplitude spectra of the $V$ data set after the subsequent steps of the deconvolution procedure. Arrows show the identified amplitude peaks of the various linear combination frequency components. {\bf a:} Residual after the contribution of the modes and their harmonics were prewhitened from the data.
{\bf b-f:} The subsequent residuals and frequency identifications, after the removal of the previously identified higher amplitude signals.}
\label{lincombs}
\end{figure}

\begin{figure}
\includegraphics[width=83mm]{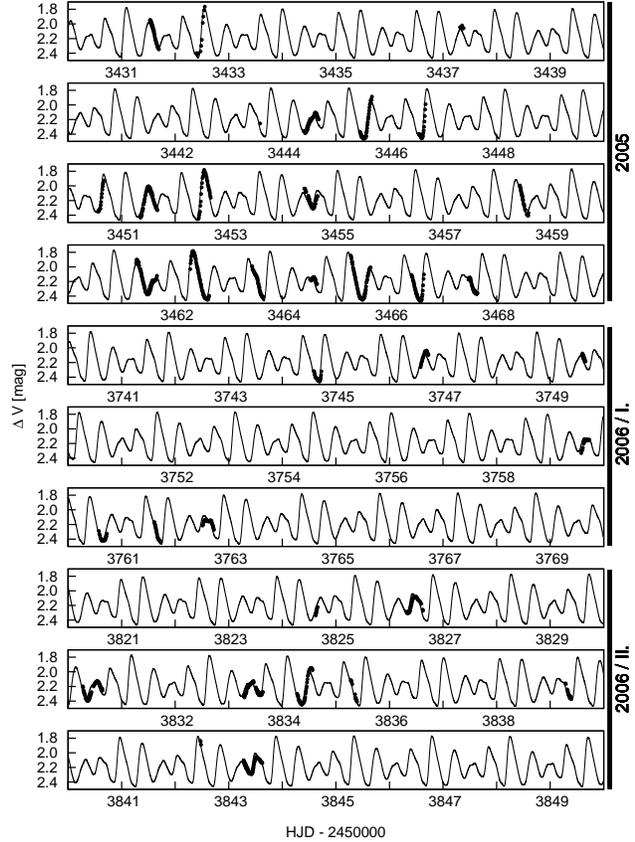}
\caption{$V$ light curve of BS Com (\emph{dots}) fitted with the Fourier solution (\emph{continous line}).}
\label{fittedv}
\end{figure}

Since the $V$ data have the highest accuracy among the different colors, we performed a standard Fourier frequency analysis (ie., discrete Fourier transformation, DFT, see e.g. Deeming 1975) on the time series of this color, using the utilities of the program package MuFrAn (Koll\'ath 1990).
Figure \ref{modes} shows the amplitude spectrum and the spectral window. The fundamental and the first overtone modes and the $\pm 1,2,...$ days aliases can be clearly distinguished. First approximation of the frequencies was obtained by a simultaneous nonlinear fit of the two modes and their harmonics. In order to find more frequency components in the spectrum, the contribution of these components were subtracted from the data. Folded $V$ light curves are shown in Figure \ref{foldedv}. Strong variations in the residual light curves clearly show that the frequencies of the two pulsation modes do not describe the light variation within the observational accuracy. Coupling frequencies are invoked in the pulsation due to strong nonlinear processes in the stellar atmosphere. In order to identify the remaining frequency components, a successive pre\-whitening procedure was performed. \sloppy{The subsequently identified frequencies were fitted simultaneously with all the previously found ones and their overall contribution was subtracted from the original data set.} This iterative method increases the signal detection probability by eliminating the strong alias structure in the vicinity of a frequency peak, enabling one to identify "hidden" low-amplitude components suppressed by aliases and separate frequency peaks that are close to each other.
The residual spectra after the subtraction of the two radial modes and their harmonics is shown in Figure \ref{lincombs}a: high amplitude linear combination components are present.
The subsequent steps of the procedure can be followed in Figures \ref{lincombs}b-f.
As a result, $15$ frequencies of the light variation were found and all but one of them were unambigously identified.\\
\indent The final set of frequencies were determined by computing numerous least squares solutions in the vicinity of the approximate values of $f_0$ and $f_1$, by fixing the harmonics ($2f_0, 3f_0, 2f_1, 3f_1, 4f_1$) and the $8$ linear combination frequencies shown in Figure \ref{lincombs} at their exact values and varying only the main frequencies. The solution yielding the smallest residual scatter was accepted. The full light curve and the corresponding Fourier fit are shown in Figure \ref{fittedv}. The final frequencies are $f_0=2.04955$ and $f_1=2.75432$. The asymptotic error estimates of the frequencies are $0.000025 \mathrm{d}^{-1}$ and $0.000020 \mathrm{d}^{-1}$ respectively. The $15$ frequency components fit the $BV(RI)_C$ data with $0.0119$, $0.0103$, $0.0099$, $0.0095$ magnitude rms scatter, respectively, which are near to the observational accuracy.

\subsection{Spectrum averaging}
In order to utilize the information given in the different colors, we employed the spectrum averaging method (SAM) as implemented by Nagy \& Kov\'acs (2006). Considering the spectrum noise as random numbers, for optimizing the noise suppressing, the spectra were weighted inversely with their variances:

{\setlength{\mathindent}{0pt}
\[
S=\frac{\sigma_V^2\sigma_R^2\sigma_I^2}{\sigma_V^2+\sigma_R^2+\sigma_I^2}\left(\frac{1}{\sigma_R^2\sigma_I^2}a_V+\frac{1}{\sigma_V^2\sigma_I^2}a_R+\frac{1}{\sigma_V^2\sigma_R^2}a_I\right)
\]
{

\noindent where $a_X$ is the amplitude spectrum of the time series in the $X$ band and $\sigma_X$ is the standard deviation around a low-order polynomial fit to the red noise of the spectra. We could made use of SAM because we have similar numbers of data points in each of the colors and the photometric sampling was also the same that had led to very similar spectral windows. If there is any signal remaining hidden in the residual with not too drastically different significance in each color, then its detection probability increases due to the decrease of the noise level. Figure \ref{sam13} shows the effect of averaging the residual spectra after all but the two smallest amplitude frequency components were subtracted from the time series. The signal at $3f_1-f_0$ slightly enhanced but there is no effect on $f_0+4f_1$ due to its very small amplitudes in $(RI)_C$. Figure \ref{sam15} shows the result of the method on the residual spectra after all the 15 frequency components were subtracted from the data. There is no additional signal over the millimagnitude level. The standard deviations of the SAM spectrum has became lower by a factor of $\sim1.5$ than the spectra of the individual colors. This decrease is lower than $\sqrt{3}$, as expected from the averaging of independent random noise, which shows that a correlated low-frequency noise is present in the spectra, due to various systematics yet unfiltered from the data.

\begin{figure}
\includegraphics[angle=-90, width=83mm]{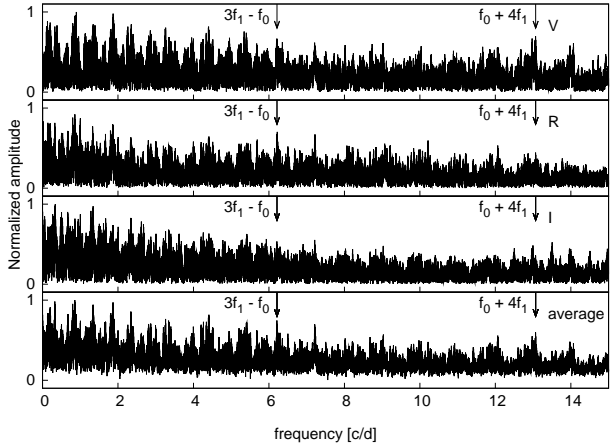}
\caption{Residual spectra after all frequencies except $f_0+4f_1$ and $3f_1-f_0$ were prewhitened from the time series. The upper three panels show the individual spectra for the $V(RI)_C$ data, the bottom panel shows their average on a normalized scale. Whereas there is no signal increase of $f_0+4f_1$ due to its very small amplitudes in $(RI)_C$, the signal of $3f_1-f_0$ seems to be enhanced.}
\label{sam13}
\end{figure}

\begin{figure}
\includegraphics[angle=-90, width=83mm]{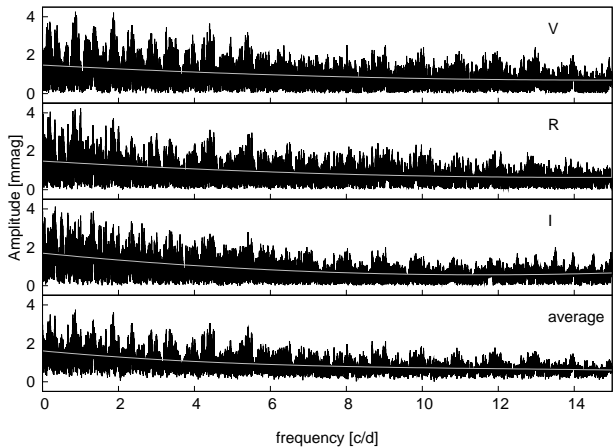}
\caption{Residual spectra after all the 15 frequencies were prewhitened from the time series. The upper three panels show amplitude spectra for the $V(RI)_C$ data, the bottom panel shows their trend-subtracted average on the same amplitude scale. The noise is suppressed by a factor of $\sim1.5$ and there is no sign of additional frequency components. A low-order polynomial fit to the red noise is shown in each panel.}
\label{sam15}
\end{figure}

\subsection{Behaviour of the Fourier-parameters}
The amplitudes of all the detected frequencies increase towards shorter wavelengths and the amplitudes of the linear combination frequency components are smaller than those of their compounding frequencies. All of the detected coupling terms are at positive linear combination frequencies of the modes except for $f_0-f_1$ and $3f_1-f_0$.
This seems to be in agreement with model results of Antonello \& Aikawa (1998).
According to their nonlinear transient double-mode Cepheid models, frequency terms of positive linear combinations have always greater amplitudes than negative linear combination ones.\\
\indent Generalized amplitude ratios are defined by
$$R_{ik}=(\vert i\vert+\vert k\vert)A_{ik} / (\vert i\vert A_{10}+\vert k\vert A_{01})$$
\noindent where $A_{ik}$ is the amplitude of the $if_0 + kf_1$ component (Jur\-csik et al. 2006). They decrease exponentially with increasing order $\vert i \vert + \vert k \vert$ (Figure \ref{rg}a), which is a similar finding as in the case of the triple-mode V823 Cas (Jurcsik et al. 2006). The generalized phase differences
$$G_{ik}=\Phi_{ik} - (i\Phi_{10} + k\Phi_{01})$$
where $\Phi_{ik}$ is the phase of the $if_0 + kf_1$ component (Antonello 1994) describe phase relations independently from the epoch. They are quasi-coherent at a given order and they show linear progression with order (Figure \ref{rg}b).

\begin{figure}
\includegraphics[angle=-90, width=83mm]{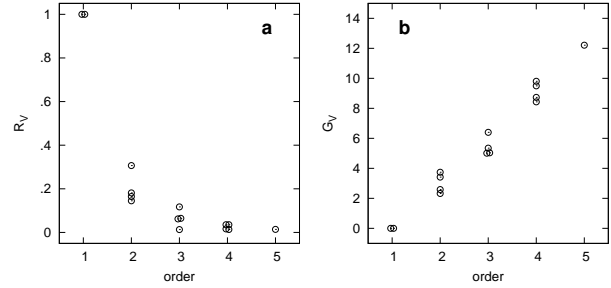}
\caption{{\bf a:} Exponential order dependence of the generalized amplitude ratios for the $V$ data set. {\bf b:} Generalized $V$ phase differences plotted against the order. The relations are very similar for the other colors.}
\label{rg}
\end{figure}

\section{Summary}
We presented preliminary results of the frequency analysis of the double-mode RR Lyrae pulsator BS Com based on a substantial set of multicolor time-series photometry. In addition to the standard single-color Fourier analysis, we combined the frequency spectra in three different colors in order to decrease the noise level. The moderate success of the spectrum averaging and the presence of low-frequency noise indicate the importance of instrumental drift and various systematics, yet to be filtered out from the time series.
The full photometric time series can be fitted with 15 frequencies within the observational accuracy. Detailed values of the derived Fourier parameters together with the photometric data and the estimated physical parameters from model results will be published elsewhere.

\acknowledgements
I would like to thank  J. Jurcsik and G. Kov\'acs for their very useful comments, and the staff of the 60 cm telescope of the Konkoly Observatory for carrying our a substantial part of the observations. The support of OTKA grant T-043504 is acknowledged.

\end{document}